\documentclass[runningheads]{llncs}
\usepackage[T1]{fontenc}
\usepackage{graphicx}
\begin{document}
\title{Regenerating Daily Routines for Young Adults with Depression through User-Led Indoor Environment Modifications Using Local Natural Materials}
\titlerunning{Regenerating Daily Routines for Young Adults}
%
\author{Ziqun Hua\inst{1,2}\and
Ao Jiang\inst{1,3,5}\textsuperscript{*}\and
Haoling Yang\inst{1,6}\and
Hao Fan\inst{4}\and
Huizhong Hu\inst{1,6}\and
Bernard Foing\inst{5}}
\authorrunning{Z. Hua et al.}
%
\institute{Nanjing University of Aeronautics and Astronautics, China \and Royal College of Arts, UK\and
Imperial College London, UK\and Southeast University, China\and International Working Group on Moon Exploration, EuroMarsMoon,ESA, Netherlands \and Jiangsu Aerospace Industrial Design and Research Institute, China}
\maketitle              
\begin{abstract}

Young adults with depression often experience prolonged indoor stays, limiting their access to natural environments and exacerbating mental health challenges. While nature therapy is recognized for its psychological benefits, existing interventions frequently require outdoor engagement, which may not be accessible for all individuals. This study explores the potential of user-led indoor modifications using local natural materials as a mental health intervention. A qualitative approach was employed to assess emotional and environmental connectedness. Participants engaged in material exploration, collection, and crafting, integrating natural elements into their living spaces. Findings indicate improved mood, increased environmental awareness, and a stronger sense of agency over personal space. The standardized intervention steps suggest the feasibility of a self-help toolkit, enabling broader implementation. This research contributes to sustainable, user-driven mental health interventions, bridging the gap between nature therapy and practical indoor applications.

\end{abstract}
\section{Introduction}
Depression among young adults aged 18–25 is an increasing concern, as prolonged indoor stays and limited outdoor engagement can exacerbate mental health conditions\cite{evans-lacko_socio-economic_2018}. While nature therapy has been shown to reduce stress and improve mood\cite{annerstedt_nature-assisted_2011,bettmann_systematic_2025}, individuals with depression often face barriers that restrict their access to outdoor environments\cite{korpela_analyzing_2014}. This highlights the need for accessible, indoor-based interventions that integrate natural elements into daily living spaces.

Research on environmental color schemes and habitability suggests that thoughtfully designed indoor environments can positively influence psychological well-being and stress reduction, particularly in confined living conditions\cite{jiang_colour_2022,stanton_habitability_2020}. Additionally, studies on color perception in environmental design indicate that specific hues affect stress response and cognitive performance\cite{jiang_young_2022,jiang_short-term_2023}. Beyond color, the incorporation of natural materials such as wood, stone, and plants into indoor spaces has been linked to reduced stress and improved well-being\cite{bringslimark_association_2008,park_ornamental_2009}.

Nature therapy research suggests that both direct and indirect exposure to natural elements fosters cognitive restoration and emotional resilience\cite{capaldi_relationship_2014,jimenez_associations_2021}. Simultaneously, behavioral activation theory\cite{jacobson_behavioral_2001} supports the idea that engaging in small, meaningful environmental modifications can foster a sense of agency and positive emotional states\cite{kennedy_behavioral_2024}. Environmental psychology further emphasizes that personalizing one’s surroundings strengthens place attachment and enhances well-being\cite{gifford_environmental_2014,nisbet_connectedness_2020}.

This study investigates whether user-led modifications of indoor spaces using local natural materials can provide psychological and environmental benefits for young adults with depression. Specifically, it examines (1) whether these modifications improve mood and well-being, (2) how incorporating natural materials affects environmental connectedness, and (3) the feasibility of a structured self-help toolkit to enable broader application.

With the growing success of self-help interventions\cite{richardson_measure_2019} and green prescriptions\cite{menhas_does_2024}, this research explores an alternative scalable, user-driven approach to mental health support. By addressing barriers to nature exposure and leveraging behavioral activation principles, this study contributes to mental health research and environmental design practices, promoting an accessible and sustainable strategy for enhancing well-being.

\section{Methods} 
\subsection{Participants}  
Eight participants (aged 18–25) with self-reported depressive symptoms were recruited through online forums and university counseling services. Eligibility was confirmed using a standardized mood scale to assess emotional states and ensure participants were experiencing depressive symptoms but were not in acute crisis.  

\subsection{Data Collection} 
Participants documented their daily routines and living environments through personal diaries and photographs to track behavioral and environmental changes. Mood fluctuations were measured using  using the Beck Depression Inventory (BDI)\cite{beck_inventory_1961}and self-reported emotion scales, during, and after the intervention. The Inclusion of Nature in Self (INS) Scale \cite{kleespies_measuring_2021} was administered pre- and post-intervention to assess changes in perceived environmental connectedness.

\subsection{Behavioral Intervention} 
Participants were encouraged to identify and collect natural materials from accessible locations within their daily environments, including nearby green spaces, household items, or overlooked natural elements. The exploration process was self-directed, allowing participants to determine suitable materials based on availability and feasibility. They were guided to observe their surroundings attentively, reconsider previously ignored materials, and engage with nature in ways adaptable to their mobility and preferences. They were encouraged to experiment with material transformations through hands-on crafting techniques, including weaving, painting, and cutting. This approach aimed to enhance environmental connectedness while fostering creativity and a sense of accomplishment\cite{keyes_creating_2024}.
  
\subsection{Data Analysis}
\begin{figure}
\includegraphics[width=\textwidth]{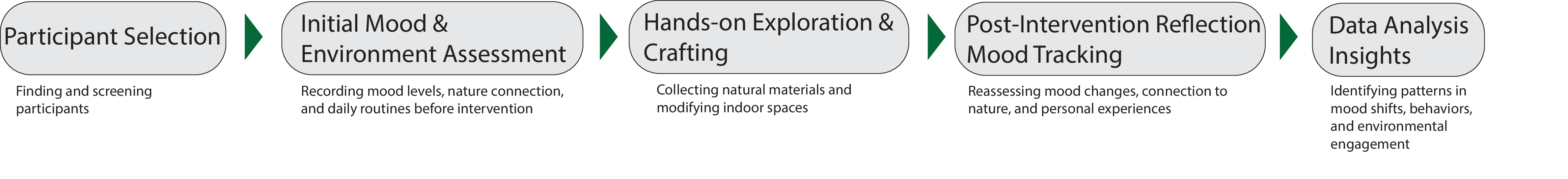}
\caption{Overview of the study workflow, illustrating participant recruitment, intervention steps, and data analysis process.} \label{figure1.eps}
\end{figure}

A thematic analysis\cite{braun_using_2006} was conducted on qualitative data from interviews, diaries, and observations, identifying patterns related to emotional responses, behavioral engagement, and environmental perception shifts. NVivo software was used to enhance consistency and reliability in coding. A visual summary of the study workflow is provided (see Fig.~\ref{figure1.eps}).

\section{Case Study: Reed Carpet Replacement}

One participant identified carpet as an essential indoor element suitable for modification. Observational and interview data indicated that selecting a modification target was an iterative process, informed by reflections on daily interactions with the home environment. During material exploration, the participant discovered a wetland park nearby and identified reeds as a sustainable, traditionally used material.

Following material analysis, weaving was selected as the preferred crafting technique due to the material’s flexibility and durability. The reeds were harvested, dried, and woven into a handmade carpet, replacing the synthetic one. Post-intervention reflections revealed psychological benefits, including a sense of accomplishment, reduced stress, and increased engagement with natural materials\cite{rhee_effects_2023}. Participants also reported heightened awareness of ecological resources and a greater sense of control over their environment\cite{nisbet_connectedness_2020}.

\begin{figure}
\includegraphics[width=\textwidth]{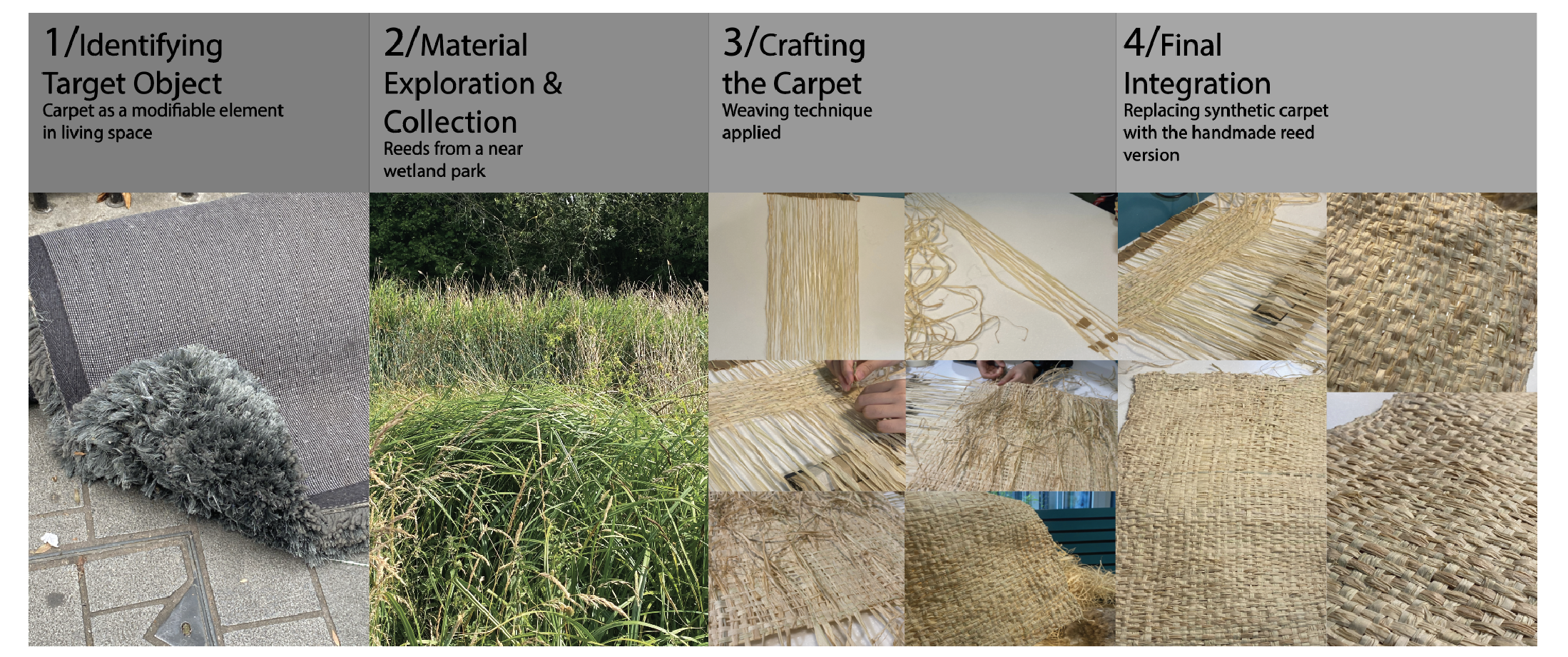}
\caption{Step-by-step process of the Reed Carpet Modification, illustrating the participant’s journey from identifying a target object to material exploration, processing, crafting, and final integration into the living space.} \label{figure2.eps}
\end{figure}

This case study exemplifies the potential for structured, self-led environmental modifications to foster both psychological well-being and ecological awareness (see Fig.~\ref{figure2.eps}).

\section{Key Findings}
The intervention demonstrated measurable psychological, environmental, and practical benefits, addressing the research questions. Participants reported improved mood, reduced passivity, and a sense of accomplishment through crafting. The structured activity provided a purposeful, calming experience, with some participants voluntarily extending their engagement beyond the study. Additionally, participants developed a stronger awareness of local natural resources, shifting from initial unfamiliarity to recognizing the value of overlooked natural elements. The presence of handcrafted natural materials in their living spaces reinforced an ongoing connection to nature\cite{zhao_effects_2023}. The standardized modification steps proved replicable and adaptable, with participants successfully following guidelines and expressing confidence in applying the approach independently. Future research should explore how digital or instructional resources could enhance broader adoption, ensuring greater accessibility and applicability across different settings.

\section{Discussion}
This study contributes to both the theoretical understanding and practical application of environmental modifications as a mental health intervention. By demonstrating the psychological and ecological benefits of user-led modifications using local natural materials, the research reinforces existing theories on nature therapy, behavioral activation, and environmental psychology while proposing an accessible, sustainable alternative to conventional interventions.

While this study highlights the feasibility and benefits of user-led indoor modifications, several limitations should be acknowledged. The small sample size limits generalizability, and further studies with diverse participant demographics are needed. Additionally, the long-term psychological effects of these modifications remain uncertain, requiring longitudinal research. While participants successfully implemented modifications, some required initial guidance, suggesting that additional instructional resources (e.g., digital toolkits or facilitated workshops) could enhance accessibility and adoption. Future work should explore the integration of digital platforms, investigate cultural variations in material selection, and assess how user-led environmental changes interact with other mental health interventions to maximize therapeutic impact.

\section{Conclusion}

This study demonstrates that user-led indoor modifications using local natural materials offer a scalable, sustainable, and psychologically beneficial approach for young adults with depression. The findings reinforce the potential of nature-integrated, hands-on interventions as a low-barrier alternative to traditional therapy.

The feasibility of a standardized self-help toolkit further highlights the potential for wider adoption and future scalability. By integrating insights from nature therapy, behavioral activation, and environmental psychology, this research contributes to the development of accessible, user-centered interventions that foster both psychological well-being and ecological engagement.


%
%
%
\bibliographystyle{splncs04}
\bibliography{hcii2025}

\end{document}